\documentclass[11pt]{article}
\usepackage{hyperref}
\usepackage{graphicx}
\usepackage{subfigure}
\pdfoutput=1
\begin{document}
\title{Wake Turbulence of Two NREL 5-MW Wind Turbines Immersed in a Neutral Atmospheric Boundary-Layer Flow }
\author{Jessica L. Bashioum, Pankaj K. Jha, Dr. Sven Schmitz \\
\\\vspace{6pt} Department of Aerospace Engineering, \\ The Pennsylvania State University, University Park, PA 16802, USA \\ \\ Dr. Earl P. N. Duque \\\vspace{6pt} Intelligent Light - Applied Research Group, \\ Rutherford, NJ 07070}

\maketitle
%% The abstract (in this file, and that submitted as text to arXiv) should
% include the exact phrase
%% "fluid dynamics video" or "fluid dynamics videos"
\begin{abstract}
%This is a sample arXiv article illustrating the use of fluid
%dynamics videos.
The fluid dynamics video considers an array of two NREL 5-MW turbines separated by seven rotor diameters in a neutral atmospheric boundary layer (ABL). The neutral atmospheric boundary-layer flow data were obtained from a precursor ABL simulation using a Large-Eddy Simulation (LES) framework within OpenFOAM. The mean wind speed at hub height is $8m/s$, and the surface roughness is $0.2m$. The actuator line method (ALM) is used to model the wind turbine blades by means of body forces added to the momentum equation. The fluid dynamics video shows the root and tip vortices emanating from the blades from various viewpoints. The vortices become unstable and break down into large-scale turbulent structures. As the wakes of the wind turbines advect further downstream, smaller-scale turbulence is generated. It is apparent that vortices generated by the blades of the downstream wind turbine break down faster due to increased turbulence levels generated by the wake of the upstream wind turbine.
\end{abstract}
% main text
\section{Introduction}
%The {\em hyperref} package is used to make links to the videos.
%%% The format is: \href{URL of video}{name that will appear in the text}
%Two sample videos are
%\href{http://ecommons.library.cornell.edu/bitstream/1813/8237/2/LIFTED_H2_EMS
%T_FUEL.mpg}{Video
%1} and
%\href{http://ecommons.library.cornell.edu/bitstream/1813/8237/4/LIFTED_H2_IEM
%_FUEL.mpg}{Video
%2}.
The wind industry faces a number of challenges today, some of which involve the aerodynamics within the wind farm. Wind turbine wakes interact with turbines located downstream, with other wakes, and with the turbulent atmospheric boundary layer (ABL). The ABL has different stability states during a diurnal cycle, thus affecting the array efficiency in a wind farm \cite{1,2}. The actuator line method (ALM) \cite{3} implemented into an ABL-LES solver created with OpenFOAM has demonstrated its potential to model large wind farms and overall wake effects \cite{4}. The present fluid dynamics video shows highly-resolved LES simulation data of two NREL 5-MW wind turbines separated by seven rotor diameters that are operating in a neutral atmospheric boundary-layer flow.
\section{Computational Methodology}
A precursor ABL simulation is performed first on a LES grid with dimensions of 3 km x 3 km x 1 km and a resolution of $15m$. For a wind speed at hub height of $8m/s$ and an assumed surface roughness of $0.2m$ over flat terrain, a quasi-stationary state is achieved after approximately $12000$ sec of real time, which can be correlated to an associated Obukhov length scale and eddy turn-over time in an atmosphere of neutral stability state. Precursor boundary-layer data are extracted at the inlet plane(s) of the LES grid at an interval of $5$ sec for about $2000$ sec after a quasi-stationary state has been achieved. These data serve as the inflow for the turbine-array simulation using the ALM. The grid has the same outer dimensions than the precursor ABL grid, however, with three layers of nested refined grids \cite{4,5} so that the grid resolution is about $2$m around the turbines and throughout the wake. The fluid dynamics video shows the downstream advection of an iso-surface of vorticity magnitude ($\omega = 0.5s^{-1}$) from various viewpoints along with contours of velocity magnitude to illustrate the breakdown process of tip vortices. Detailed information on the solver settings can be found in Jha et al. \cite{5}. \newline\\
\begin{figure}[ht]
  \centerline{\includegraphics[height=6.3cm]{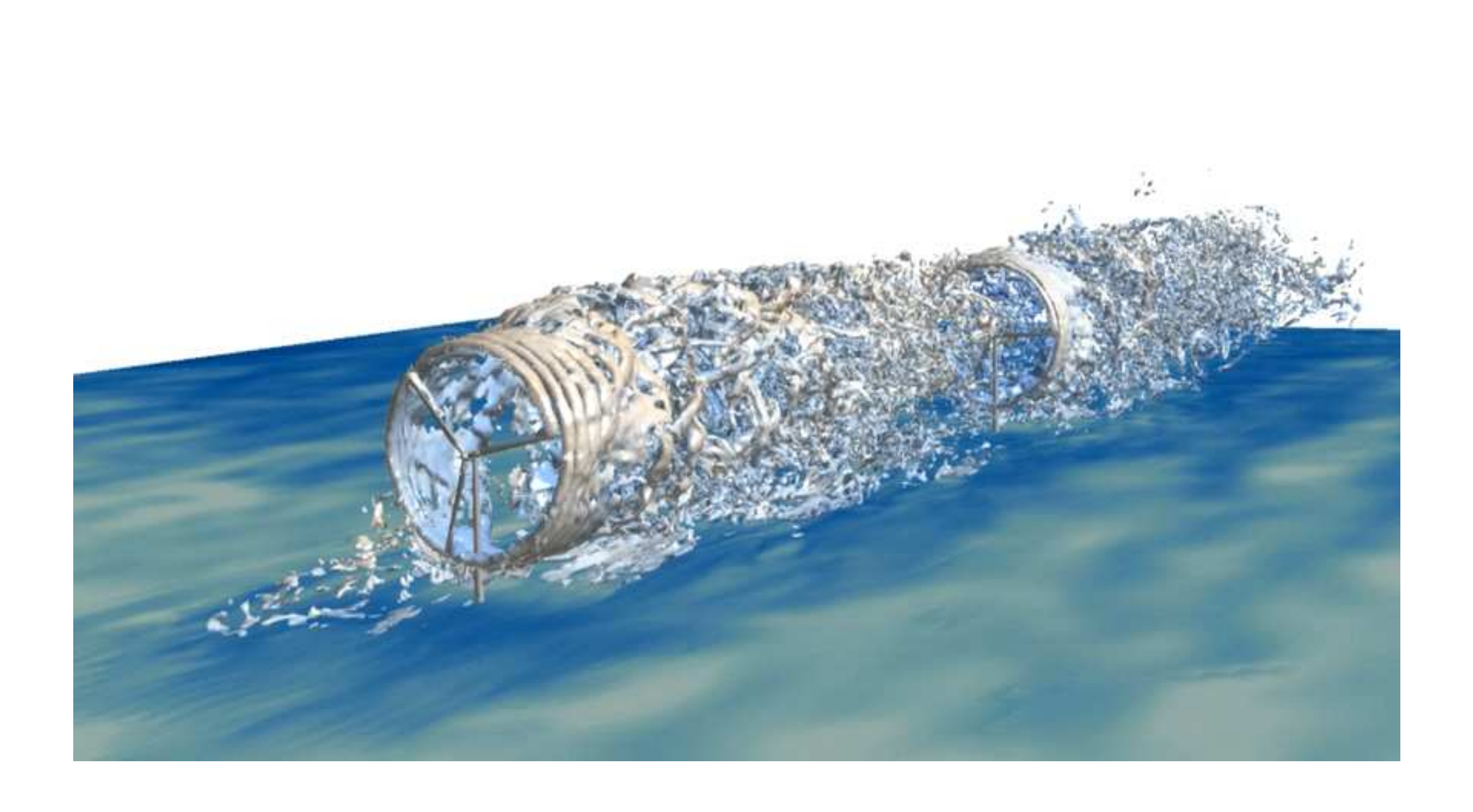}}
  \caption{Array of two NREL 5-MW wind turbines in neutral ABL flow (Iso-surface of vorticity magnitude, $\omega = 0.5s^{-1}$).}
\label{fig:1}
\end{figure}
The fluid dynamics video was created making use of a data extract workflow; also known as a "XDB workflow". In this workflow, iso-surface and coordinate cut-plane data extracts were created using the FieldView CFD post-processing tool \cite{6} in batch mode. Iso-surfaces for various levels of vorticity magnitude were computed. The coordinate cut planes were created for several directions and in an axial plane slicing the rotor tower position.
Computed OpenFoam data were saved to disk for $0.5$ to up to $2$ rotor revolutions. After each solver run, six (6) concurrent FieldView batch jobs were launched to create the surface extracts for up to 10 time steps each. Each individual batch job used a maximum of $16$ shared memory processors and up to 32GB of memory to create one extract file, "xdb file", for each time step processed. Each surface extract has the same grid resolution as the original volume grid, however, it is several orders of magnitude smaller than the original OpenFOAM volume data. For the iso-surfaces, the scalar quantities of velocity and vorticity magnitudes were saved. For the coordinate cut planes, the velocity and vorticity vectors along with Reynolds stresses were saved for future analyses.
Following, all created data extracts were transferred to a local workstation where the frames for the fluid dynamics video were rendered using FieldView in interactive mode. This process allowed for multiple viewpoints and key frame animations without having to read the volume data and recompute the iso- and coordinate surfaces.

\section{Results and Discussion}
\begin{figure}[ht]
  \centerline{\includegraphics[height=5.3cm]{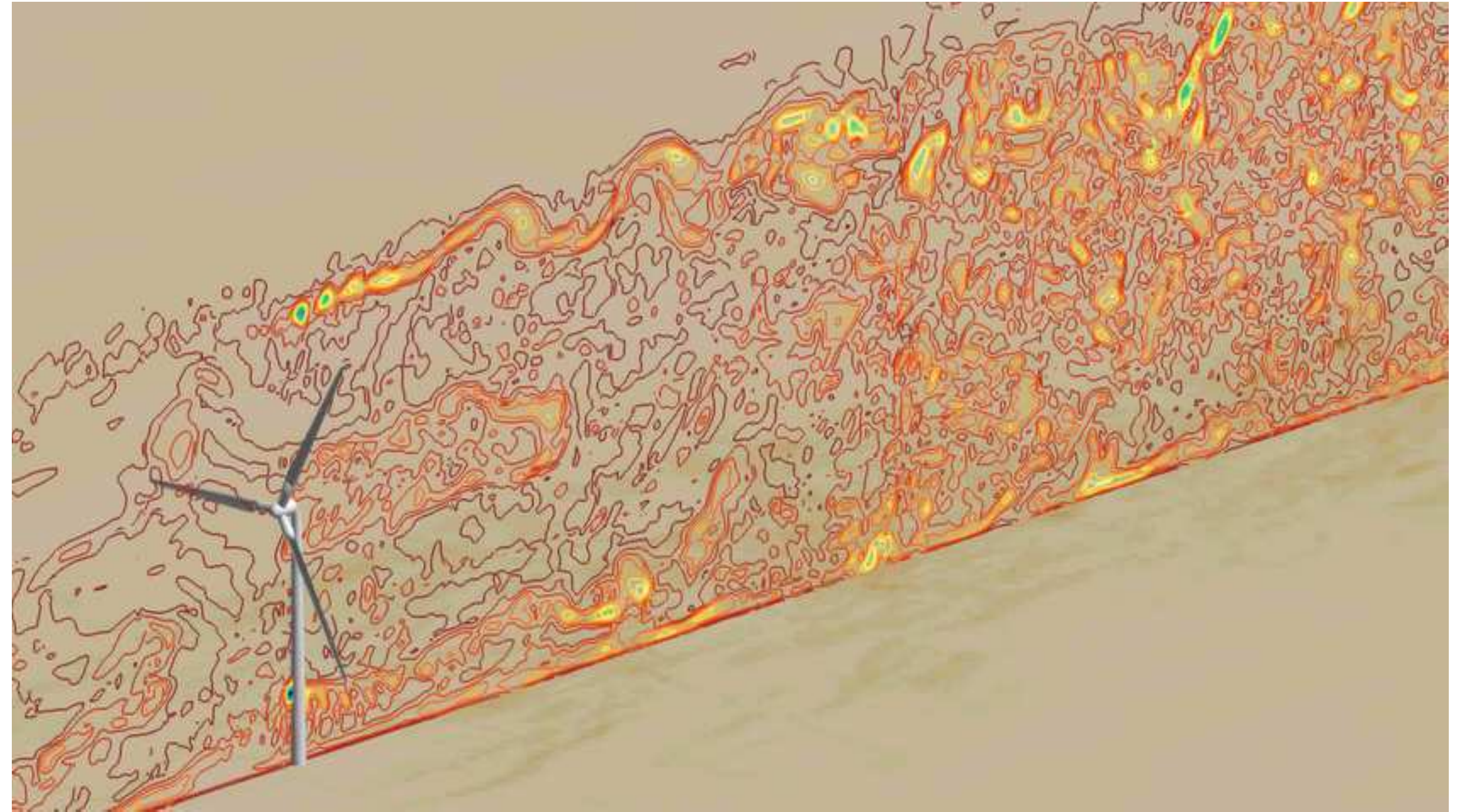}}
  \caption{Tip vortex instability of upstream NREL 5-MW wind turbine (Contours of velocity magnitude).}
\label{fig:2}
\end{figure}
\begin{figure}[ht]
\centering
\subfigure[Upstream wind turbine]{%
\includegraphics[height=4.4cm]{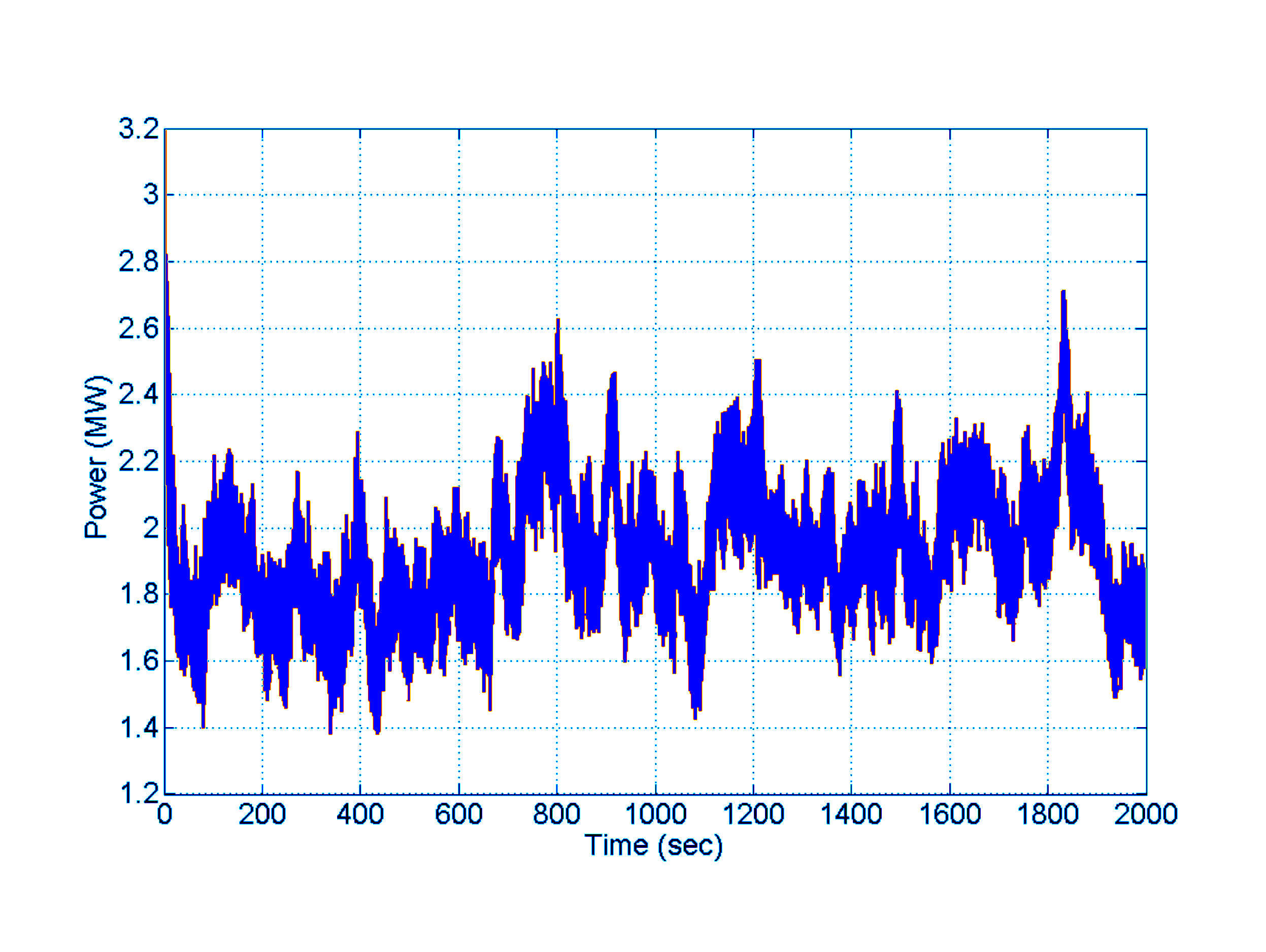}
\label{fig:3a}}
\quad
\subfigure[Downstream wind turbine]{%
\includegraphics[height=4.4cm]{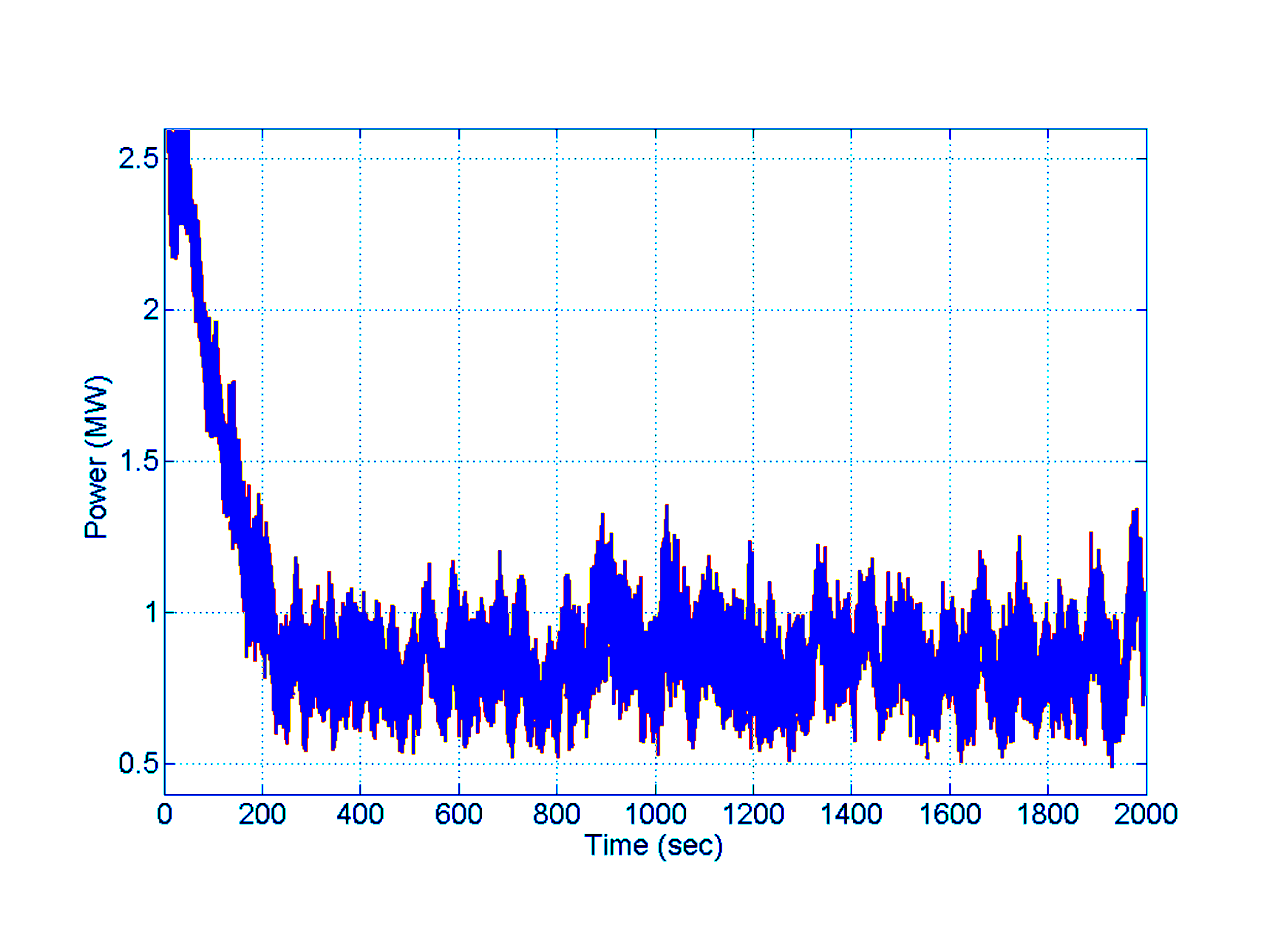}
\label{fig:3b}}
\caption{Power histories of upstream and downstream wind turbines.}
\label{fig:3}
\end{figure}

It is known to the wind energy community that wind turbines located in the wake of upstream turbines show a significant decrease in power production and an increase in blade fatigue loads. A viewpoint of the fluid dynamics video is shown in Fig. \ref{fig:1}. It can be seen that the tip vortices of the upstream turbine become unstable after about $1.5$ rotor diameters downstream. This is the so-called near-wake region in which the wake expands as expected from classical momentum theory and seen in the side view of the fluid dynamics video. Furthermore, the pairing and breakdown processes of the tip vortices can be seen in the fluid dynamics video in the mid-wake region. It is also apparent that the larger-scale turbulent structures break down into smaller-scale turbulence with distance from the upstream wind turbine. Figure \ref{fig:2} illustrates this process by contours of velocity magnitude. \newline \\
The downstream wind turbine is subject to increased turbulence levels generated by the wake of the upstream wind turbine. As a consequence of this, tip vortices emanating from the downstream wind turbine break down earlier in the wake compared to those of the upstream turbine, see Fig. \ref{fig:1}. Figure \ref{fig:3b} shows indeed that the power generated by the downstream turbine has reduced significantly compared to that of the upstream wind turbine in Fig. \ref{fig:3a}. This is associated with the fact that the momentum deficit of the upstream wind turbine has not fully recovered when it interacts with the downstream wind turbine.

\section{Conclusion}
The present fluid dynamics video shows an array of two NREL-5MW wind turbines immersed in a neutral ABL flow. Fundamental flow-physical processes such as tip vortex pairing and breakdown can be seen along with the generation of additional small-scale turbulence in the wake of a large wind turbine. Further investigations of mean and fluctuating components as well as Reynolds stresses will enhance the physics-based understanding of wind turbine wakes. \newline \\
This work was supported by the Department of Energy Grant DE-EE0005481. The fluid dynamics video would not have been possible without the generous supply of \emph{Fieldview 14} software licenses by Intelligent Light and the continuous support of Earl P. N. Duque from the Applied Research Group at Intelligent Light.

%\section*{References}

\end{document}